\documentclass[a4paper]{article}
\usepackage[pdftex]{graphicx} 
\usepackage[english]{babel}
\usepackage[bf,small,nooneline]{caption}
\usepackage[round]{natbib}
\title{Mars as a comet: Solar wind interaction on a large scale} 
\begin{document}
\date{October 8, 2015}
\author{Mats Holmstr{\"o}m\thanks{Swedish Institute of Space Physics, PO~Box~812, SE-98128~Kiruna, Sweden. (\texttt{matsh@irf.se})}, and Xiao-Dong Wang$^*$}
\maketitle

\begin{abstract}
Looking at the Mars-solar wind interaction on a larger spatial 
scale than the near Mars region, the planet can be seen as an 
ion source interacting with the solar wind, in many ways like a comet, 
but with a smaller ion source region. 
Here we study the interaction between Mars and the solar wind 
using a hybrid model (particle ions and fluid electrons). 
We find that the solar wind is disturbed by Mars out to 
100 Mars radii downstream of the planet, and beyond. 
On this large scale it is clear that the escaping ions can be 
classified into two different populations. 
A polar plume of ions picked-up by the solar wind, and a more 
fluid outflow of ions behind the planet. 
The outflow increases linear with the production up to 
levels of observed outflow rates, then the escape levels off for 
higher production rates. 
\end{abstract}

\newcommand{\dfrac}[2]{\frac{\displaystyle#1}{\displaystyle#2}}
\newcommand{\pder}[2]{\frac{\displaystyle\partial#1}{\displaystyle\partial#2}}

\section{Introduction}
The interaction between the solar wind and Mars 
provides one way that the atmosphere of Mars can be lost. 
As soon as neutrals in Mars atmosphere are ionized, e.g., by photoionization, 
they are subject to electro-magnetic forces and can be 
accelerated and subsequently escape the planet. 
To quantify the present loss of atmosphere at Mars, it is therefore 
of interest to understand how ions can escape the planet under 
different conditions. 
In the past, ion outflow has been observed by the Phobos-2~\citep{Lundin89} 
and the Mars Express~\citep{Lundin04} missions. 
In the most general sense, the loss of ions from Mars can be seen as 
a mass loading~\citep{Szego00} of the solar wind, 
similar in some ways to the mass loading at comets, 
but with a smaller ion source region. 
If we go to even larger spatial scales, the fluxes of ions produced 
near Mars will be too tenuous to affect the solar wind, and behave 
like an expanding cloud of ions.  
Here we focus on the intermediate region, far away from the planet, 
but not so far away that the solar wind is unperturbed. 
What is the morphology of the ion outflow, and the global interaction 
of the planet with the solar wind?  
What is the morphology of the bow shock in the far wake region? 
The tool we use is a hybrid model (particle ions and fluid electrons) 
of the interaction between Mars and the solar wind, that 
we describe in the next section. 

The solar wind interaction with Mars has been modeled by many 
groups using different kinds of models, e.g., 
empirical~\citep{Kallio97}
test particles~\citep{Fang08}, 
magnetohydrodynamic (MHD)~\citep{Ma04}, multi-fluid~\citep{Harnett06} 
and hybrid models~\citep{Brecht90,Kallio01,Modolo05,Bosswetter07}. 

The advantage of fluid models is that they are computationally less 
expensive than hybrid models, allowing finer grid resolution. 
On the other hand, hybrid models contains a more accurate 
description of the ion physics, something important at Mars where 
the planet is of similar size as the ion length scales, 
such as the ion gyro radius and the ion inertial length. 
We can note that a comparison of many different models of the 
interaction of Mars with the solar wind found significant 
differences in the general interaction for the different models~\citep{Brain10}. 

One thing in common to past model investigations is that they have 
consider the near Mars region, up to a few planet radii away from the planet.  
This is natural since the 
aim of the studies has been focused on topics such as 
determining the loss of heavy ions to space. 
This requires that the model resolves the ionospheric 
region well with high spatial resolution.  
This implies a high computational cost 
for a large simulation domain, so the domain is chosen just large 
enough for the process under study.  For ion escape studies 
the simulation domain just has to be large enough that there 
is no significant returning heavy ion flux. 

Previous studies has found that the escaping ions can be 
classified into two different populations. 
A polar plume of ions picked-up by the solar wind, and a more 
fluid outflow of ions behind the planet. 
In hybrid simulations~\citet{Brecht12} found tailward electric fields 
in the hemisphere opposite to the solar wind convective electric field 
the accelerate heavy ions such that they escape downstream behind the planet. 
The ion plume has been described based on 
Mars Express observations~\citep{Liemohn14}, and has also been studied using 
test particle simulations~\citep{Curry13}.

\section{Model}
In what follows we describe the details of the computer model we use to study 
the large scale plasma interaction between Mars and the solar wind. 
The model has previous been applied to the solar wind interactions of 
the Moon~\citep{EPS12} and the plasma interactions of 
Callisto~\citep{Jesper15}. 
First we describe the plasma model, then the heavy ion production, 
and finally the boundary conditions. 

\subsection{Hybrid model}
In the hybrid approximation, ions are treated as particles, 
and electrons as a massless fluid. 
In what follows we use SI units. 
The trajectory of an ion, $\mathbf{r}(t)$ and $\mathbf{v}(t)$, 
with charge $q$ and mass $m$, is computed from the Lorentz force, 
\[
  \dfrac{d\mathbf{r}}{dt} = \mathbf{v}, \quad
  \dfrac{d\mathbf{v}}{dt} = \dfrac{q}{m} \left( 
    \mathbf{E}+\mathbf{v}\times\mathbf{B} \right), 
\]
where $\mathbf{E}=\mathbf{E}(\mathbf{r},t)$ is the electric field, 
and $\mathbf{B}=\mathbf{B}(\mathbf{r},t)$ is the magnetic field.  
From now on we do not write out the dependence on $\mathbf{r}$ and $t$. 
The electric field is given by 
\begin{equation}
  \mathbf{E} = \dfrac{1}{\rho_I} \left( -\mathbf{J}_I\times\mathbf{B} 
  +\mu_0^{-1}\left(\nabla\times\mathbf{B}\right) \times \mathbf{B} 
   - \nabla p_e \right) + \frac{\eta}{\mu_0} \nabla\times\mathbf{B}, 
\label{eq:E}
\end{equation}
where $\rho_I$ is the ion charge density, 
$\mathbf{J}_I$ is the ion current density, 
$p_e$ is the electron pressure, 
$\eta$ is the resistivity, and
$\mu_0=4\pi\cdot10^{-7}$ is the magnetic constant. Then 
Faraday's law is used to advance the magnetic field in time, 
        \begin{equation}
          \pder{\mathbf{B}}{t} = -\nabla\times\mathbf{E}. \label{eq:F}
        \end{equation}
Note that the unknowns are the position and velocity of the ions, and 
the magnetic field on a grid, \emph{not} the electric field, since it 
can always be computed from~(\ref{eq:E}). 
Further details on the hybrid model used here, and the discretization, 
can be found in~\citep{Enumath09,Astronum10,EPS12}. 

In the wake of obstacles to the solar wind flow, e.g., behind Mars, 
the plasma densities can be low, or even zero. 
In such regions of low ion charge density, $\rho_I$, the hybrid method can have 
numerical problems.  We see from~(\ref{eq:E}) that the electric field 
computation involves a division by $\rho_I$, so regions of low density will 
have large electric fields. 
This can lead to numerical instabilities, due to large gradients 
in the electric field, and due to large accelerations of ions. 
The numerical solution quickly becomes unstable. 
Here we handle regions of low ion charge density by solving a 
magnetic diffusion equation in such regions. 
A magnetic diffusion equation is also solved inside the 
obstacle to the solar wind flow, in this case inside the spherical 
inner boundary. 
See~\citet{Astronum12} for details of this approach. 
%A previous application was to study the interaction between 
%the Moon and the solar wind~\citep{EPS12}. 
%%% Shahab ref that uses the diffusion. 

\subsection{Ion production}
% See flash-mhd-planet.tex   Also figures from that???
Instead of building a complete ionospheric model with many species, 
sources, and loss terms, we only specify a source of photoions given by 
a standard Chapman production function~\citep{KandR}. 
\begin{equation}
  p(h,\chi) = p_0 e^{1-y-\sec\chi e^{-y}}, \qquad y = (h-h_0)/H, 
  \qquad h \geq 0, \quad 0 \leq\chi<\pi/2, 
\end{equation}
where $h$~[m] is the height above the planet surface, $\chi$~[rad] 
is the solar zenith angle (SZA), $p_0$~[m$^{-3}$s$^{-1}$] 
is the maximum production along the sub-solar line, at height $h_0$~[m], 
and $H$~[m] is the scale height. 
In this study we only consider atomic oxygen, O$^+$, ions, and we call them 
heavy ions (heavies) as opposed to the other species in the 
simulation, solar wind protons, H$^+$, 
and solar wind alpha particles, He$^{++}$. 
This is a very simplified ionospheric model. 
It only contains one heavy ion specie, 
has an unrealistic ion production profile, and 
does not contain any ion chemical reactions. 

%%% Fig ion production?  chapman.pdf  syms.txt 

The justification for such a simplified ionospheric model is that 
we are interested in the effects on large spatial scales, 
where we can view Mars as a source of ions in the solar wind and 
the exact details of the distribution of heavy ions close to the 
planet should not be very important.  

After selecting $h_0$ and $H$ we can then choose different $p_0$ that 
will result in different outflow rates of heavy ions. 
Since the outflow rate of heavy ions has been observed to be 
on the order of $10^{25}$~s$^{-1}$~\citep{Ramstad13}
we choose a production $p_0$ that gives an outflow on that 
order of magnitude.

\subsection{Boundary conditions}
Heavy ions will be lost from the simulation in two ways. 
Either through the inner spherical boundary, or through the 
outer boundary of the simulation box. 
Such ions are removed from the simulation.  

One interesting feature of the simulations is that the whole 
magnetosphere becomes unstable after some time if the simulation cells are 
larger than about 1000~km.  The bow shock then flares out and 
can hit the upstream boundary of the simulation domain.  

\subsection{Parameter values}
Unless otherwise noted, the parameter values used in the simulation 
runs are as follows. 

The coordinate system is centered at Mars, with 
the $x$-axis directed toward the Sun, so the solar wind 
flows opposite to the $x$-axis.  The radius of Mars, $R_M=3380$~km. 

We have used typical solar wind conditions at Mars from~\citet{Brain10}. 
In the upstream solar wind the interplanetary magnetic field 
(IMF) is $(-1.634,2.516,0.0)$~nT with a magnitude of 3~nT. 
% 57 degrees to the negative x-axis 
This direction of the IMF is along the nominal Parker spiral at Mars. 
The solar wind speed is 400~km/s, 
with a proton number density of 3~cm$^{-3}$. 
We also include He$^{++}$ with a number density of 0.06 cm$^{-3}$. 
The solar wind proton temperature is $5\cdot10^{4}$~K, 
and the electron temperature is $1.2\cdot10^{5}$~K. 
For these plasma parameters in the solar wind, 
the ion inertial length is 131~km, 
the Alfv\'{e}n velocity is 38~km/s, %the ion thermal velocity is 29~km/s, 
the thermal proton gyro radius is 100~km (the gyro time is 22~s), 
and the ion plasma beta is 0.6. 

The maximum ionospheric production rate of O$^+$, 
$p_0=10^4$~s$^{-1}$m$^{-3}$ at a height of $h_0=500$~km. 
The scale height, $H=250$~km.  
Note that the scale and peak heights of the ionosphere are much larger than in 
reality.  However, this does not affect the results much since the cell size 
of the simulations are larger than these heights.  Thus, the ionosphere 
is not resolved at all in the simulations.  This is confirmed by the fact 
that changing the scale height to $H=150$~km did not change the results much. 
The weight of the heavy ion macroparticles are $4\cdot 10^{20}$ 
(this many O$^+$ ions per macroparticle. 
We keep the produced number of heavy ion macroparticles constant. 
So if $p_0$ is increased by a factor, then the weights of the 
heavy ion macroparticles are increased by the same factor. 
The simulation box +$x$ face is an inflow boundary and -$x$ an outflow 
boundary.  The boundary conditions on the other faces are periodic. 
The simulation domain is divided into a Cartesian grid with 
cubic cells of size 700~km.  
The extent of the simulation 
box is $[-186.6,15]\times[-100.8,100.8]\times[-100.8,100.8] \cdot 10^6$ km
$\approx [-55.2,4.44]\times[-29.8,29.8]^2 R_M$. 
At the inflow boundary we have three H$^+$ macro particles per cell, 
and one He$^{++}$ macro particle. 
The total number of simulation macroparticles is about 100 million.  
Five subcycles of the cyclic leapfrog (CL) algorithm~\citep{Enumath09} 
are used here when updating the magnetic field, 
and the execution time is about 24 hours on 432 CPU cores.

\section{Results}
\begin{figure}
\begin{center}
  \includegraphics[width=1.0\columnwidth]{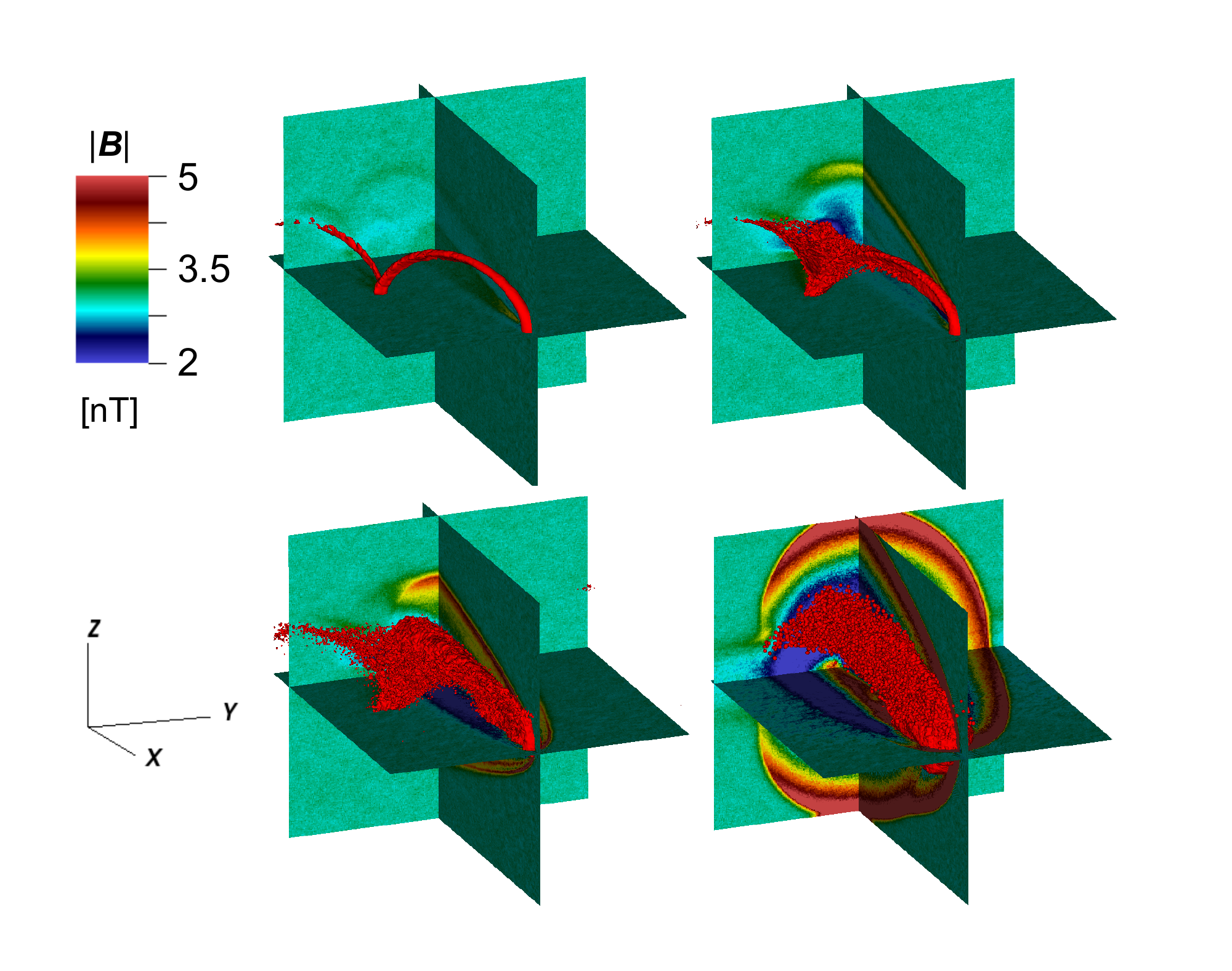}
\end{center}
\caption{  % mars-127, 126, 125, 128
Results for relative heavy ion production of 0.01 (top left), 0.1 (top right), 
1 (bottom left), and 10 (bottom right). 
Iso surface of heavy ion number density of 0.01~cm$^{-3}$ is shown in red. 
The colors of the cuts show magnetic field strength 
according to the color bar.  
The simulation domain extends to 50~$R_M$ down stream of Mars, 
and the solution is at $t=1200$~s. 
The cut perpendicular to the $x$-axis is at $x=-1.6\cdot10^{8}$~m. 
} \label{fig:channels} 
\end{figure}

\begin{figure}
\begin{center}
  \includegraphics[width=0.8\columnwidth]{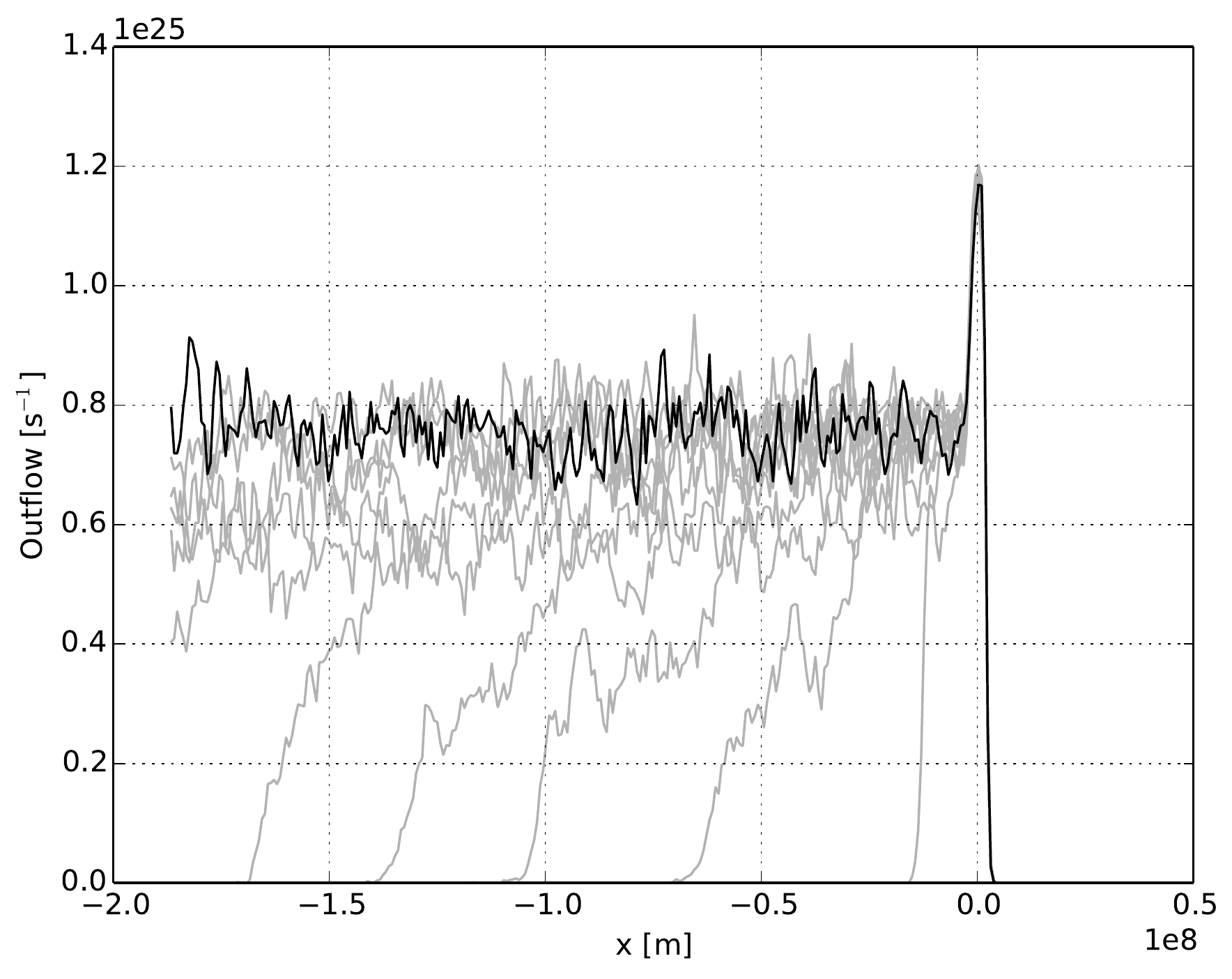}
\end{center}
\caption{  
Total O$^+$ number flux through planes perpendicular to the $x$-axis for a 
relative heavy ion production of 1, at different times. 
The black line is at the final time of 1100~s, and the grey lines are at earlier times in intervals of 100~s. 
} \label{fig:outflow} 
\end{figure}

We now investigate the results for four different relative heavy 
ion productions; 0.01, 0.1, 1, and 10. 
The morphology of the heavy ion outflow and the magnetic field 
magnitude is shown in Figure~\ref{fig:channels}. 
It is seen that the heavy ion outflow follows two different channels. 
A pick-up like escape where the ions gyrate in the interplanetary 
magnetic field (IMF) and a more fluid escape downstream of the planet. 
At the lowest production of 0.01 all the heavy ions are picked-up. 
They are still present in the higher production cases (0.1 and 1) 
but then there is also escape behind Mars. 
For the highest production of 10, all ions escape fluid like 
in the tail behind Mars.  However, the simulation had not reached a 
steady state yet for the highest production. 
%%% Fig escape channels
For the magnetic field, we can see in Figure~\ref{fig:channels} that for 
the lowest production of 0.01 no significant bow shock is formed, there 
are however field disturbances associated with the picked-up ions. 
For larger production a bow shock is formed that strengthen with increased 
production.  We also see that the outflowing ions generate field 
disturbances downstream that resembles a break in the bow shock, 

For the production 1 case we show in Figure~\ref{fig:outflow} 
the total O$^+$ number flux as a function of distance downstream of Mars, 
for different simulation times.  
We see that the simulation reaches a steady state situation after 
900~seconds. 
%%% Fig outflow vs time
% $ ./outflow_rev.py /mnt/WD/lrun/mars-125/flash_hdf5_plt_cnt_00*

\begin{figure}
\begin{center}
  \includegraphics[width=0.7\columnwidth]{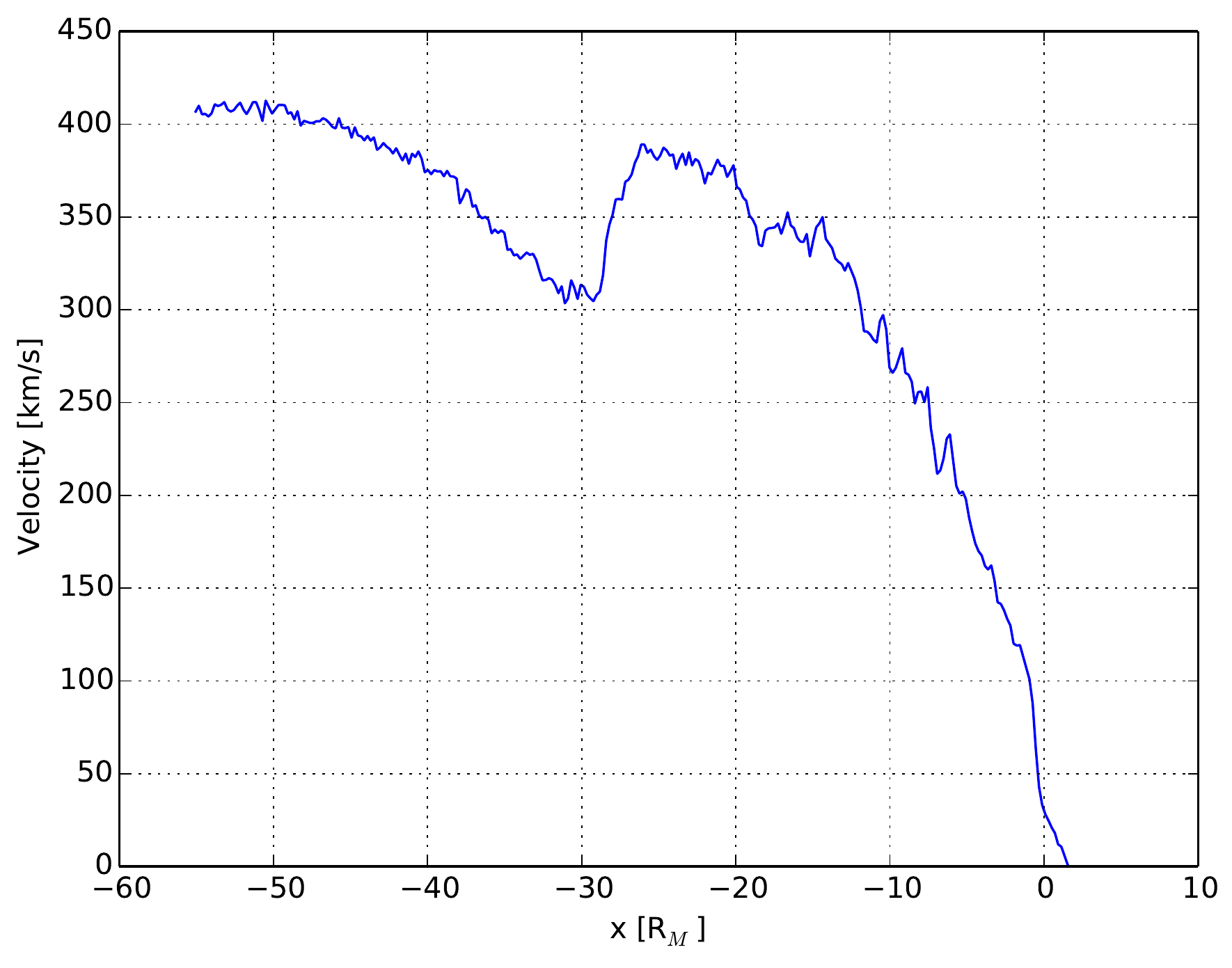}
\end{center}
\caption{  
The average velocity of O$^+$ ions in planes perpendicular to the $x$-axis 
for a relative heavy ion production of~1, after 1100~seconds. 
The velocity is computed as the total ion current divided by total charge 
in each plane. 
} \label{fig:velocity} 
\end{figure}

In Figure~\ref{fig:velocity} we show the average velocity of O$^+$ ions 
as a function of distance downstream of Mars. 
It takes 45~Mars radii before the heavy ions reach the solar 
wind velocity of 400~km/s.  Visible is also the gyration of some of the ions, 
as a drop in velocity at about -30~$R_M$.  This is due to the cycloid motion 
of the picked-up heavy ions. 
%%% Fig velocity
% matsh@pic:~/shared/projects/flash/hdf5/Flash_Visualization$ ./velocity.py /mnt/WD/lrun/mars-125/flash_hdf5_plt_cnt_0011 

We now examine how the outflow changes with production. 
In Table~\ref{tab:oflow} we show the total outflow of O$^+$ ions for different normalized production values.  We do not include the outflow for a production of 
10 since that simulation did not reach a steady state after 1200~s.  
The heavy ions accumulated over time close to the planet 
and were not transported away to escape. 
The outflow is determined as 
an average behind Mars at a time of steady state, as shown in 
Figure~\ref{fig:outflow} for the case when the production is 1.  
\begin{table}
  \caption[]{Total outflow of O$^+$ ions as a function of normalized production.}
  \label{tab:oflow}
  \begin{center}
  \begin{tabular}{p{0.2\linewidth}|l|l|l}
  Production        & 0.01   & 0.1    & 1      \\ \hline
  Outflow [s$^{-1}$] & 2.8$\cdot 10^{23}$ & 2.6$\cdot 10^{24}$ 
                                                 & 0.7$\cdot 10^{25}$ \\
  \end{tabular}
  \end{center}
\end{table}
We see in Table~\ref{tab:oflow} that the outflow increases linear with 
the production for low productions (from 0.01 to 0.1). 
For higher productions (from 0.1 to 1) the outflow levels off. 

\begin{figure}
\begin{center}
  \includegraphics[width=1.0\columnwidth]{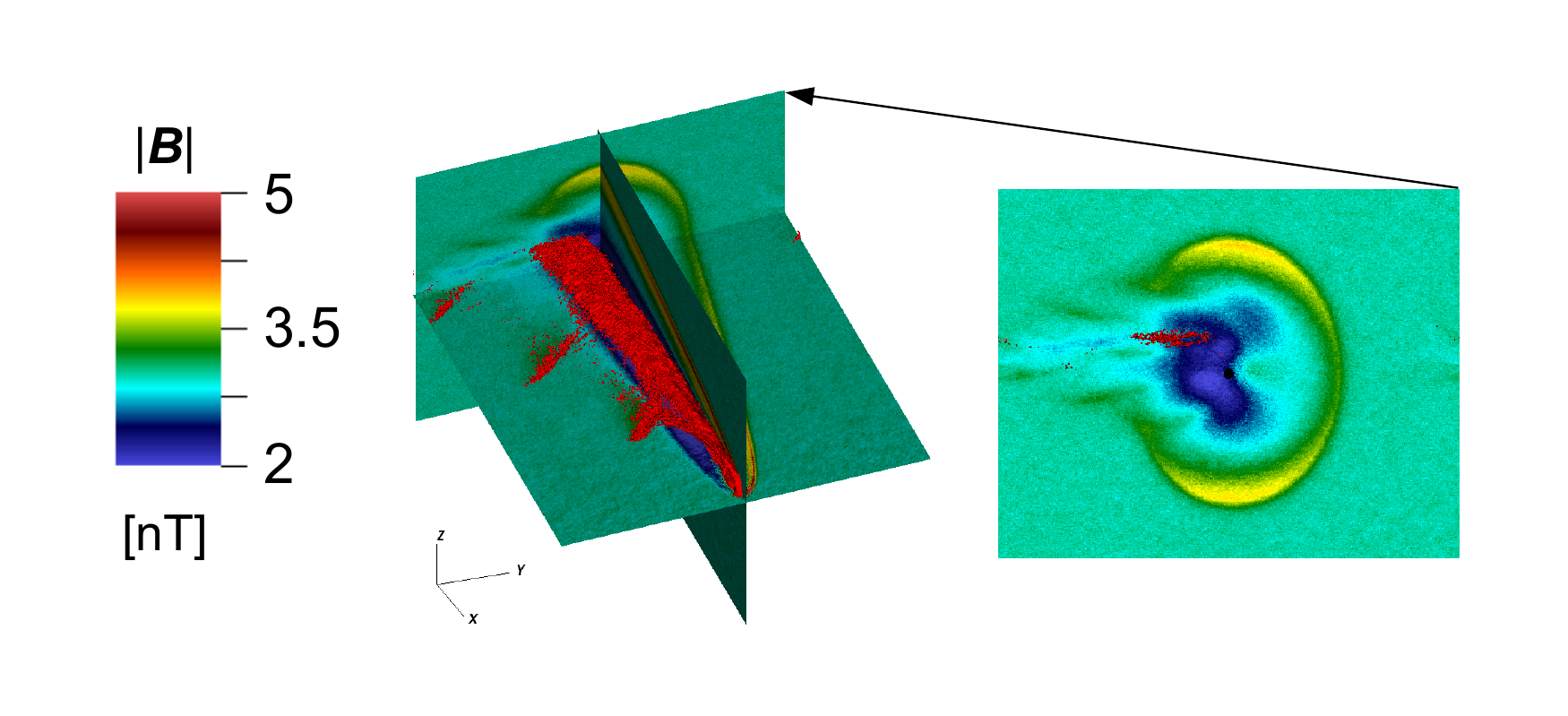}
\end{center}
\caption{  % mars-125c2
Results for a relative heavy ion production of 1. 
Iso surface of heavy ion number density of 0.02~cm$^{-3}$ is shown in red. 
The colors of the cuts show magnetic field strength 
according to the color bar.  
The simulation domain extends to 100~$R_M$ down stream of Mars, 
and the solution is at $t=1200$~s. 
The cut perpendicular to the $x$-axis is at $x=-3.4\cdot10^{8}$~m, 
and the black disk at the center is the projection of Mars onto this plane. 
} \label{fig:100} 
\end{figure}

%%% Fig 100 RM with cut
In Fig.~\ref{fig:100} we show the magnetic field strength and the 
heavy ions for a run out to 100~$R_M$ downstream of Mars. 
On this large scale, 
we clearly see the mix of two different populations of escaping O$^+$. 
One population is the picked-up O$^+$ that is visible to the left, 
moving in cycloid motion perpendicular to the IMF. 
The other population is a more fluid-like bulk flow of 
heavy ions going down stream in the $-x$-direction. 
We also see how the Mach cone is disturbed and 
penetrated by the escaping heavy ions. 
The bulk flow is several $R_M$ above the plane of the IMF, and is 
not extended much even at this far distance from the planet. 
There is some spreading in the plane of the IMF (the $xy$-plane), 
but not much in the direction perpendicular to the IMF plane. 

\begin{figure}
\begin{center}
  \includegraphics[width=1.0\columnwidth]{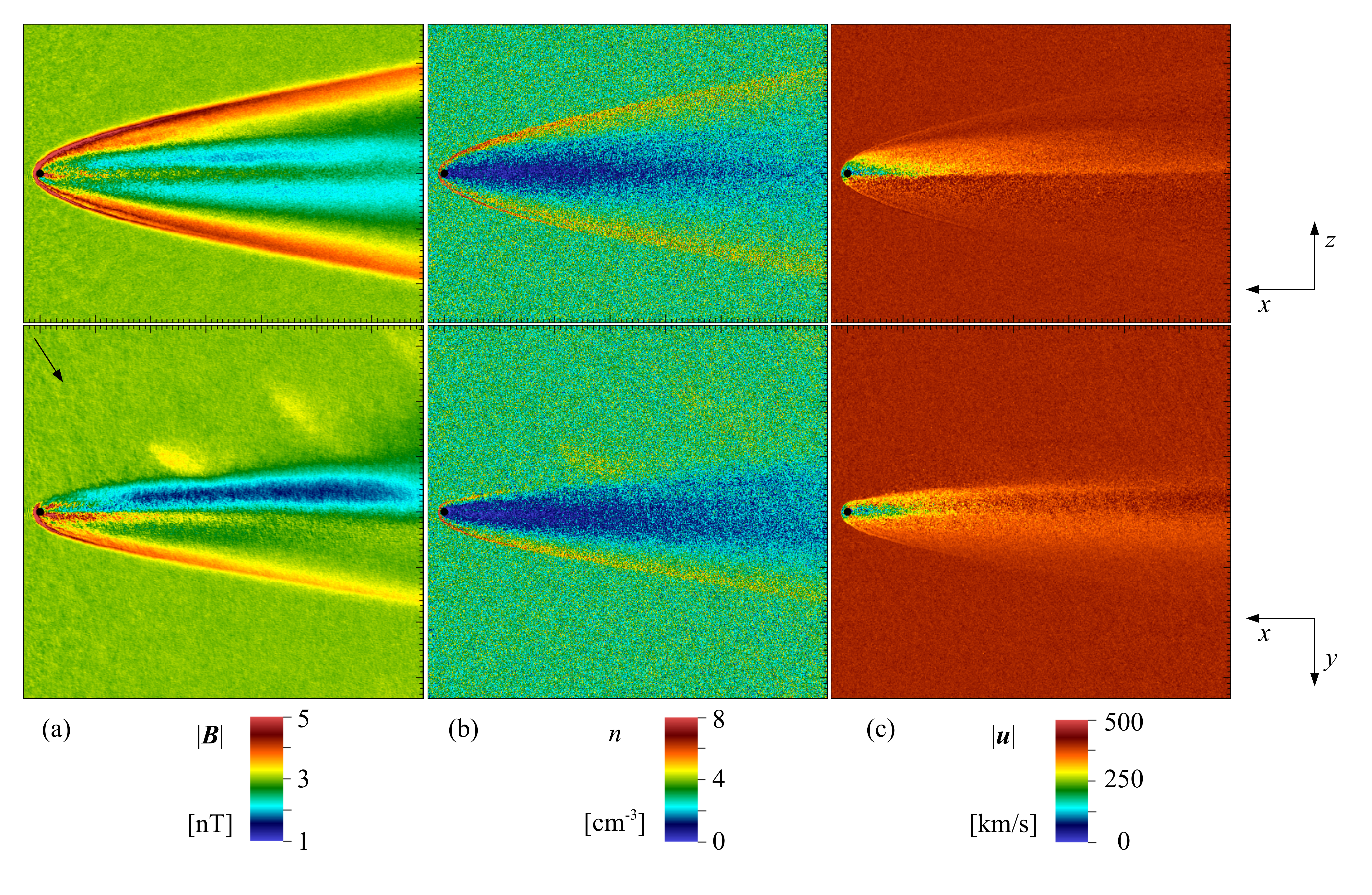}
\end{center}
\caption{  % mars-125c2
Magnetic field magnitude (a), H$^+$ number density (b), 
and H$^+$ velocity magnitude (c), 
for a relative heavy ion production of 1.
Cuts in the $xz$-plane (top) and in the $xy$-plane (bottom). 
The direction of the interplanetary magnetic field (IMF) 
in the $xy$-plane is indicated by a black arrow.  
The simulation domain extends to 100~$R_M$ down stream of Mars, 
and the solution is at $t=1200$~s. 
The black disk is the projection of Mars. 
The maximum magnetic field magnitude is 9.3~nT and 
the maximum number density is 15~cm$^{-3}$. 
} \label{fig:cuts} 
\end{figure}

For the simulation out to 100~$R_M$ downstream of Mars, we also 
show the magnetic field magnitude and the proton number density 
in Fig.~\ref{fig:cuts}.  
%%% Mag field plot w cuts + H+ cut ?
On this scale the bow shock is non-existent in the plane of the IMF 
on the side where the bow shock normal would be quasi parallel to the IMF. 
There are however disturbances associated with the gyrating heavy ions. 
We can also see that the bow shock is split on the quasi perpendicular side, 
as has been observed closer to Mars in earlier simulations~\citep{Modolo05}. 
This is seen also in the proton number density, 
along with also disturbances associated with the gyrating heavy ions. 
The protons are slowed down close to the planet and in the wake, but are 
accelerated to solar wind speeds further down stream.  

\section{Conclusions}
We have studied the interaction of Mars with the solar wind on a 
large scale, out to 100 Mars radii downstream of the planet. 
A simplified ionosphere as a source of O$^+$ ions was used, since 
details of the ionosphere should not be important on these scales. 
It was found that on this large scale it is easy to see that the 
heavy ions escape along two different channels, the proportion 
depending on the production of heavy ions. 
For small production the heavies escape mostly as pick-up ions, 
while for large production the heavies escape as a bulk flow. 
The ions undergoing pick-up type escape will gyrate around the IMF 
and leave imprints in the magnetic field and proton densities. 
The ions subject to a fluid like escape will flow downstream the 
wake behind Mars without much spreading even at these far distances. 

The outflow increases linearly with production up to 
levels of observed outflow rates, then the escape levels off for 
higher production rates.  
It is an open question if there is a maximum escape rate. 

The pick-up process will also separate ions with different mass per charge. 
Here we have only studied O$^+$, but the main Martian ionospheric species 
O$^+$, O$_2^+$, and CO$_2^+$ will all follow different trajectories. 
This would lead to three different cycloid escape channels. 

We found that the solar wind is disturbed by Mars out to 
100~$R_M$ downstream of the planet, and beyond. 
We can note that,  
as seen from Earth, 100~$R_M$ is of similar size as the full moon. 

\section*{Acknowledgements}
This work was supported by The Swedish Research Council, grant 621-2011-5256, 
and was conducted using resources provided by the Swedish National Infrastructure for Computing (SNIC) at the High Performance Computing Center North (HPC2N), Ume\aa\ University, Sweden.
The software used in this work was in part developed by the 
DOE NNSA-ASC OASCR Flash Center at the University of Chicago. 

%\bibliography{mars}

\end{document}